\documentclass[fleqn,twoside]{article}
\usepackage{espcrc2}

\usepackage{graphicx}
\usepackage{epsf}
\usepackage[figuresright]{rotating}
\usepackage{amssymb}

    \newcommand{\tr}{\mbox{Tr}}

\newcommand{\AmS}{{\protect\the\textfont2
  A\kern-.1667em\lower.5ex\hbox{M}\kern-.125emS}}

\title{\vspace{-3.65cm}
       {\normalsize DESY 01-168}    \\[-0.2cm]
       {\normalsize October 2001}   \\
       \vspace{2.72cm}
Glueball and gluelump spectrum in abelian projected QCD \thanks{Talk given by V. Bornyakov.}}

      \author{V. Bornyakov\address{NIC/DESY Zeuthen, Platanenallee 6, 15738 Zeuthen, Germany  \\ },%
        \thanks{On leave of absence from IHEP, Protvino, Russia}
        G. Schierholz $^{\rm a,}$
\address{Deutsches Elektronen-Synchrotron DESY,
                    D-22603 Hamburg, Germany},
and T. Streuer \address{Sektion Physik, Universit\"at M\"unchen, 80333 
M\"unchen,  Germany }
}
 
\begin{document}

\begin{abstract}
We study glueball and gluelump spectra calculated after abelian projection in
both quenched and $N_f=2$ full QCD. The abelian projection is made after MA gauge
fixing. We demonstrate that both spectra can be recovered despite the problem
with positivity. We suggest the interpretation of some of the gluelump states
in the language of the abelian projected theory.
\vspace{1pc}
\end{abstract}

\maketitle

\section{INTRODUCTION}

The effective infrared theory obtained
after abelian projection must reproduce the low mass hadron spectrum
at least in qualitative agreement with the real spectrum.
To check this in the maximally abelian (MA) projection \cite{klsw}
we calculate glueball and gluelump spectrum in the abelian projected
(AP) SU(3) theory. We also discuss results for QCD with dynamical 
fermions.
The glueball spectrum in AP $SU(2)$ theory has been studied in 
\cite{stack} and good agreement with the $SU(2)$ spectrum has been found.
The low hadron masses have been computed in AP $SU(3)$ \cite{suzuki}. 
This study, though being limited in precision, allows to draw conclusion
about qualitative agreement with the spectrum of the unprojected 
theory. 
There is another motivation of our work. It has been claimed \cite{dp}
that since abelian projection breaks $SU(3)$ invariance (even global)
the 'new hadronic' states must appear in a theory which are absent in the
experimental spectrum. We will suggest a solution of this problem.

\noindent
\section{SIMULATION DETAILS} 
We follow the standard procedure of abelian projection for $SU(3)$
in MA gauge  \cite{klsw}.
To fix MA gauge a simulated annealing algorithm \cite{bbms} has been 
employed with the aim to reduce effects of Gribov copies. 
We generated one gauge copy per configuration. 
Our computations have been done on a $16^3\cdot 32$ lattice. 
In the  quenched case we used $O(400)$ configurations at $\beta=6.0$.
For the $N_f=2$  full QCD $O(300)$ configurations at $\beta=5.29, 
\kappa=0.1350$ generated by QCDSF \cite{hs} have been used. These 
two sets have roughly equal lattice spacing with $r_0/a \approx 5.3$.

\section{{GLUEBALLS IN QUENCHED/FULL AP QCD}}

\noindent Let us introduce lattice gauge field $U_{x,\mu} \in SU(3)$ and
abelian projected field 
$u_{x,\mu}=\mbox{diag} \{ e^{i\theta_{x,\mu}^1},e^{i\theta_{x,\mu}^2},e^{i\theta_{x,\mu}^3} \} \in U(1) \times U(1)$.
The glueball correlator has a general form 
\vspace{-.05cm}
\begin{equation}
 \Gamma(t)= < \tr G(0) \tr G^{\dagger}(t)> - <\tr G>^2 \label{glb}\,
\end{equation}
\vspace{-.05cm}
The  zero momentum operator 
\vspace{-.05cm}
\begin{equation}
 G(t)=\sum_{\vec{x}}G(\vec{x},t)\, 
\end{equation}
\vspace{-.5cm}
\[ G(\vec{x},t)=\sum_C \left(U(C_x) \pm U^\dagger(C_x)\right), \ U(C)=\prod_{l \in C} U_l \label{glb_op}\, \]
\vspace{-.05cm}
belongs to one of the three representations of the cubic group: 
$\mbox{A}^{++},\mbox{E}^{++},\mbox{T1}^{+-} $. In (\ref{glb_op})
closed paths $C$ and sign are chosen properly to get particular 
representation.
To obtain the corresponding projected correlator we make the
following substitution in (\ref{glb_op}):
 $\ U(C) \rightarrow \ u(C) $.
We use square loops of a size up to $8\times8$ and 
apply smearing or fuzzing to $u_{x,i}$.
Finally, the effective mass is extracted as usual:
\begin{equation}
a\mbox{m}_{\mbox{eff}}(t) = -\mbox{log} \left [ \frac{\Gamma(t+1)}{\Gamma(t)} \right ] 
\end{equation}
In Fig.~1 we present our results for projected $\mbox{m}_{\mbox{eff}}(t)$ for
$0^{++}$ glueball. The straight lines show central value and error
bars of the $0^{++}$ glueball mass obtained at this $\beta$ value in
\cite{mt} in the unprojected theory.
\begin{figure}[t!]
\vspace{.2cm}
\hbox{
\epsfysize=5.0cm
\epsfxsize=7.cm
\epsfbox{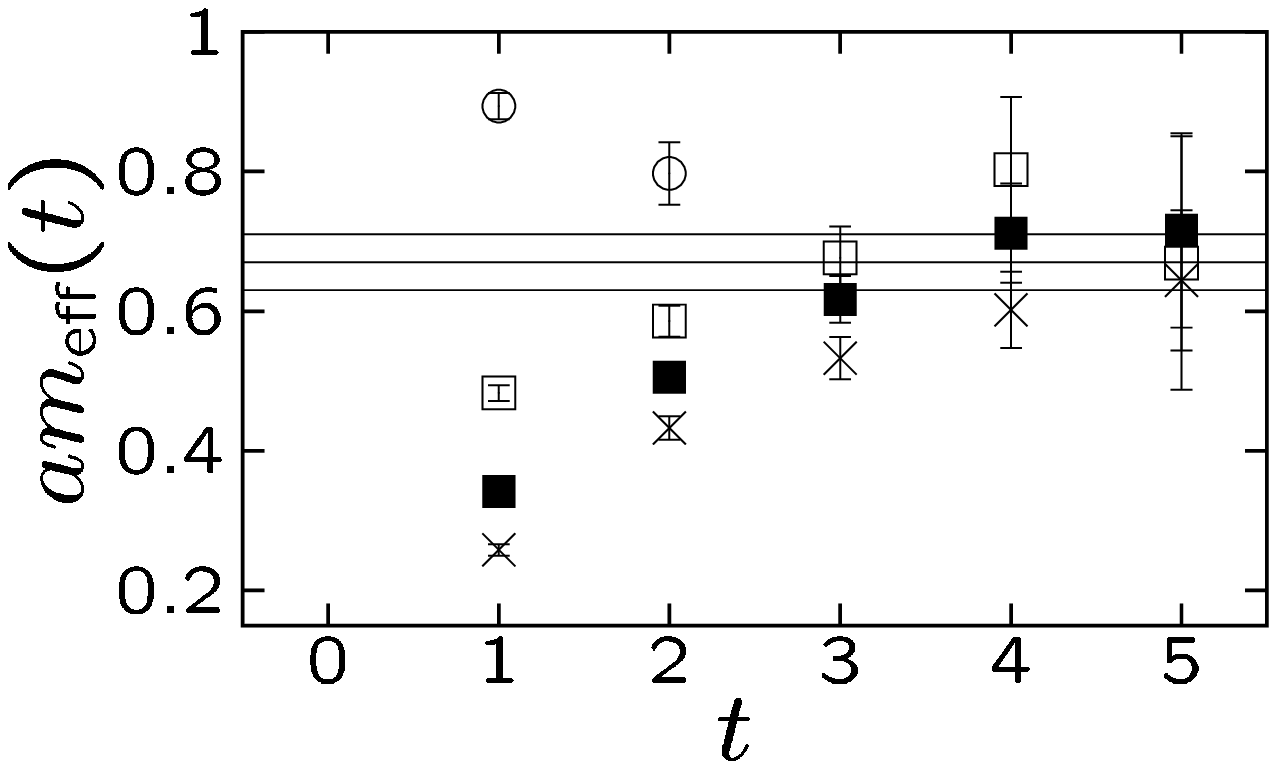}}
\vspace{-.8cm}
\caption{\label{0++} {\it $0^{++}$ glueball effective mass for the 
following operators: 
$1\times 1$ unsmeared $(${\Large$\circ$}$)$, $2\times 2$ smeared 
$(\square)$, $4\times 4$ smeared $(\blacksquare)$, $3$ levels of 
fuzzing $(\times)$.
\vspace{-.4cm}
}} 
\end{figure}
One can see unusual behaviour of $\mbox{m}_{\mbox{eff}}(t)$ for
smeared/fuzzed operators. This is due to lack of reflection 
positivity for gauge noninvariant operators used in our computations, 
as has been discussed in \cite{stack}. We found consistency between 
our results
obtained with various operators which assures us that our computation
of the glueball masses is meaningful.
Good agreement with results of \cite{mt} can be seen from Fig.~1. 
Similar agreement was found for $2^{++}$ and $1^{+-}$ glueballs. 
We repeated
 our computation on full QCD configurations and obtained results 
consistent within error bars with results for the quenched case.
\section{{GLUELUMPS IN QUENCHED/FULL AP QCD}}
\noindent
The gluelump was introduced  in \cite{cm1}.  
It is not a physical particle. 
It can be seen as a glueball with one gluon infinitely massive
or as a static adjoint quark with color screened by dynamical gluon 
field.
Its lowest energy determines a scale where adjoint string should break. 
The gluelump correlator has the following form:
\begin{equation}
 \Gamma(t) =\mbox{Tr}\left [ G(\vec{x},0) \lambda_a \right ] \ S_{ab}(\vec{x},t) \ \mbox{Tr} \left [ G(\vec{x},t) \lambda_b \right ] 
\end{equation}
\[ S(\vec{x},t) = \prod_{\tau=1}^{t} U^{adj}_{(\vec{x},\tau),0} \]
\noindent
The gluelump spectrum has been computed both in $SU(2)$ \cite{cm2}
and in $SU(3)$ \cite{cm3}.
It is worth noting that $1^{+-}$ and $1^{--}$ gluelump correlators 
coincide with the gauge invariant correlators of the magnetic and
electric fields $<B^a_i(x)S_{ab}(x,y)B^b_i(y)>$, 
$<E^a_i(x)S_{ab}(x,y)E^b_i(y)>$ 
\cite{simonov}, where $S_{ab}$ is a Schwinger line.
In the stochastic vacuum model \cite{ds} the field strength
correlator has exponential decay and determines the gluon correlation 
length $T_g$ \cite{ds}.
Within the scope of this  model $T_g$ is a fundamental parameter.
The field strength correlator has been computed on the lattice.
The lattice results for $T_g$ can be summarized as follows: in $SU(2)$
$T_g = 0.15 \div 0.2$ fm \cite{dg1_ith}, in $SU(3)$ $T_g = 0.1 \div 0.2$ fm \cite{dg2_ith}. 
The ground state energy of the gluelump, $E_g$, is related to $T_g$
\cite{simonov}:
\begin{equation}
 \frac{1}{T_g} = m_g, \ m_g = E_g - \mbox{divergent selfenergy} 
\end{equation}
The abelian projected gluelump correlator is:
\[ \tilde{\Gamma}(t) = \sum_{\alpha=3,8}\mbox{Tr}\left( \tilde{G}(\vec{x},0) \lambda_\alpha \right) \mbox{Tr}\left( \tilde{G}(\vec{x},t) \lambda_\alpha \right) \]
Note the absence of the Schwinger line in the projected correlator.
This implies that the energy computed from this correlator has
no divergent part. 
Recently a connection between the field strength correlator and the 
dual photon propagator of an effective infrared QCD (dual Abelian 
Higgs model) has been suggested \cite{bbdv}. If this is true, then 
the dual photon mass is equal to $m_g$ introduced above. In this sense
we can say that the dual photon corresponds to the $1^{+-}$ gluelump
and thus is not observable.
We computed correlators $\tilde{\Gamma}(t)$ for $1^{+-}, 0^{++}, 2^{++}$
gluelumps using the same techniques as in the computation of the projected
glueball spectrum. 
\begin{figure}[t!bh]
\vspace{.2cm}
\hbox{
\epsfysize=4.1cm
\epsfxsize=7.cm
\epsfbox{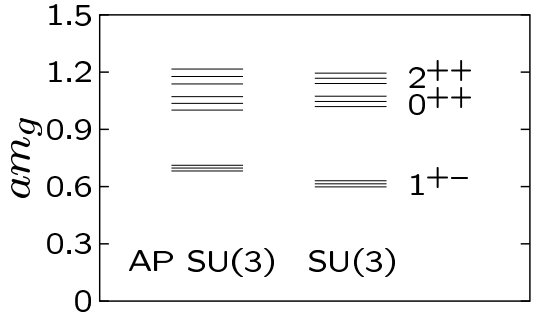}}
\vspace{-.8cm}
\caption{\label{comp}\it{Comparison of the gluelump spectra obtained 
with and without abelian projection.
\vspace{-.4cm}
 }}
\end{figure}
To compare with results of \cite{cm3} we
need to subtract the divergent part from these results. This we make
in the way applied to $SU(2)$ gluelumps in \cite{cm2}:
$ m_{g} = E_{g} - \frac{1}{2} V_0^{adj} $, where $V_0^{adj}$ is 
a selfenergy of the static adjoint quark-antiquark pair. To estimate 
$V_0^{adj}$ we use the relation $V_0^{adj}=\frac{9}{4}V_0^{fund}$, which
holds according to results of \cite{bali}, and take the value 
$V_0^{fund}=0.63(2)$ from \cite{bs}. 
Our results presented in Fig.~2 show good qualitative agreement with
the $SU(3)$ spectrum of gluelumps obtained in \cite{cm3}. 
Similar to the glueballs case the gluelump spectrum in AP QCD agrees 
well with results depicted in  Fig.~2.\\
It is easy now to suggest the interpretation of the 'new hadron' states 
considered in \cite{dp}, namely the states generated by operators 
$\bar{\psi}(x) \lambda_{3,8} \psi(x)$. They are abelian projections of 
adjoint-mesons \cite{cm3} determined by the correlator
\vspace{-.1cm}
\begin{equation}
 \Gamma(t) = \tr\left [ G_a(\vec{x},0) \right ] S_{ab}(\vec{x},t) \tr \left [ G_b(\vec{x},t) \right ] 
\vspace{-.1cm}
\end{equation}
\[ G_a(x) = \bar{\psi}(x) \lambda_a \psi(x) \]
\vspace{-.1cm}
After abelian projection the correlator is: 
\begin{equation}
 \tilde{\Gamma}(t) = \sum_{\alpha=3,8}\tr\left [ G_\alpha(\vec{x},0) \right ] \tr\left [ G_\alpha(\vec{x},t) \right ] 
\end{equation}
\vspace{-.1cm}
The adjoint-baryons and their abelian projection can be constructed 
analogously.
Note that we always use Weyl invariant operators since our gauge
fixing does not break Weyl invariance.

\section{{CONCLUSIONS}}

Our results for glueball and gluelump spectra in AP $SU(3)$ show 
good qualitative agreement with corresponding results obtained in
$SU(3)$. To make conclusion about how good agreement is quantitatively 
the extrapolation to the continuum limit would be necessary. Similar 
computations made in AP full QCD at $m_\pi/m_\rho=0.76$
have not revealed any essential changes in the spectrum.
We  proposed  solution of the problem
of 'new hadrons', raised in \cite{dp}: these states are abelian 
counterparts of gluelumps (or adjoint mesons/baryons) 
and thus cannot be detected in experiments.

\noindent
This work is partially supported by INTAS grant 00-00111.

\end{document}